\documentclass[twocolumn]{openjournal}
\usepackage{natbib}

\usepackage{graphicx,amsmath,amssymb,amstext}
\usepackage{amsbsy,amsfonts,amsthm,color}
\usepackage[colorlinks,linkcolor=blue,citecolor=blue,urlcolor=blue ]{hyperref}
\usepackage[utf8]{inputenc}
\usepackage{float}
\usepackage{lipsum}
\usepackage{verbatim}
\usepackage[normalem]{ulem}
\usepackage{orcidlink}
\usepackage{soul}
\usepackage{subcaption}

\defcitealias{Prole2023}{LP23}
\defcitealias{Schauer2021}{AS21}

\begin{document}

\title{Fast X-ray Transient Detection with AXIS: Application to Magnetar Giant Flares\\[-5.ex]}

\author{Michela Negro$^{*1}$\orcidlink{0000-0002-6548-5622}}
\author{Zorawar Wadiasingh$^{2,3,4}$\orcidlink{0000-0002-9249-0515}} 
\author{George Younes$^{4,5,6}$\orcidlink{0000-0002-7991-028X}}
\author{Eric Burns$^{1}$\orcidlink{0000-0002-2942-3379}}
\author{Anirudh Patel$^{7}$\orcidlink{0009-0000-1335-4412}}
\author{Brian D. Metzger$^{7,8}$ \orcidlink{0000-0002-4670-7509}}
\author{Todd A. Thompson$^{9,10,11}$ \orcidlink{0000-0002-1730-1016}}
\author{Daryl Haggard$^{12, 13}$ \orcidlink{0000-0001-6803-2138}}
\author{S. Bradley Cenko$^{2}$ \orcidlink{0000-0003-1673-970X}}

\affiliation{$^{1}$Department of Physics \& Astronomy, Louisiana State University, Baton Rouge, LA 70803, USA}
\affiliation{$^{2}$NASA Goddard Space Flight Center, 8800 Greenbelt Road, Greenbelt, MD 20771, USA}
\affiliation{$^{3}$Department of Astronomy, University of Maryland, College Park, MD 20742, USA}
\affiliation{$^{4}$Center for Research and Exploration in Space Science and Technology, NASA/GSFC, Greenbelt, MD 20771, USA}

\affiliation{$^{4}$Center for Space Sciences and Technology, University of Maryland, Baltimore County, Baltimore, MD 21250}
\affiliation{$^{5}$Astrophysics Science Division, NASA Goddard Space Flight Center, Greenbelt, MD 20771, USA}
\affiliation{$^{6}$Center for Research and Exploration in Space Science and Technology, NASA/GSFC, Greenbelt, MD 20771, USA}

\affiliation{$^{7}$Department of Physics and Columbia Astrophysics Laboratory, Columbia University, New York, NY 10027, USA}
\affiliation{$^{8}$Center for Computational Astrophysics, Flatiron Institute, 162 5th Ave, New York, NY 10010, USA}

\affiliation{$^{9}$Department of Astronomy, Ohio State University, 140 West 18th Avenue, Columbus, OH 43210, USA}
\affiliation{$^{10}$Center for Cosmology \& Astro-Particle Physics, Ohio State University, 191 West Woodruff Ave., Columbus, OH 43210, USA}
\affiliation{$^{11}$Department of Physics, Ohio State University, 191 West Woodruff Ave., Columbus, OH 43210, USA}

\affiliation{$^{12}$Department of Physics, McGill University, 3600 rue University,
Montr\'eal, Quebec City H3A 2T8, Canada}
\affiliation{$^{13}$McGill Space Institute, McGill University, 3550 rue University, Montr\'eal, Quebec City H3A 2A7, Canada}

\email{$^*$email: michelanegro@lsu.edu}

\begin{abstract}
Magnetar giant flares (MGFs) are among the most luminous high-energy transients in the local universe, consisting of a short, intense MeV $\gamma$-ray spike followed by a softer, pulsating X-ray tail and possibly delayed radioactive emission. While only three Galactic events have been firmly detected, several extragalactic candidates have recently been reported, motivating the need for sensitive, rapid-response $\gamma$- and X-ray facilities to constrain their rates and energetics. We present a feasibility study of detecting MGFs with the Advanced X-ray Imaging Satellite (AXIS), focusing on two complementary pathways: (i) serendipitous discovery of the prompt $\gamma$-ray spike within the field of view, and (ii) rapid follow-up of MGF tails in nearby galaxies. Using sensitivity rescaling and volumetric rate estimates, we find that serendipitous detection of prompt spikes during the mission lifetime is possible but unlikely, primarily because of their short duration and primarily because of their short duration and hard spectrum, in the assumption that the hard gamma-ray spectrum can be reliably extrapolated to the instrument’s energy range. In contrast, AXIS’s superior sensitivity, if accompanied by fast repointing capabilities, offer an extraordinary opportunity to detect pulsating X-ray tails out to $\sim$20 Mpc, enabling the first extragalactic measurements of periodic modulations from a magnetar and potentially constraining emission geometry and fireball physics. Finally, we evaluate the detectability of soft X-ray line emission from r-process nucleosynthesis in MGFs, finding that such signals are extremely faint and confining the detection to Galactic distances. Our study offer a general framework for assessing the detectability of short transients with future missions.
\end{abstract}


\section{Introduction}
Extragalactic fast X-ray transients (FXTs) are a diverse class of short-lived, high-energy phenomena characterized by rapid flares in the X-ray band, typically lasting from fractions of a second to a few hours. They are detected sporadically and often without clear counterparts at other wavelengths, making their physical origins difficult to pinpoint. The transient nature and wide range of timescales suggest multiple underlying populations: some events may be related to prompt gamma-ray burst (GRB) emission, GRB afterglows, tidal disruption events, active galactic nuclei activity, or exotic compact object activity \citep{2017MNRAS.467.4841B,2022A&A...663A.168Q,2023A&A...675A..44Q}. Because of their unpredictability and rarity, FXTs remain one of the most intriguing topics in time-domain astrophysics, highlighting the importance of wide-field monitoring instruments and coordinated multiwavelength follow-up.

Among the possible engines powering fast X-ray transients (FXTs) are magnetars. These are the most highly magnetized neutron stars, with surface magnetic fields inferred to exceed $10^{14}$ Gauss from timing measurements \citep{kouveliotou1998x,kouveliotou1998discovery}, consistent with theoretical expectations \citep{1995MNRAS.275..255T} and subsequent observational evidence across the X-ray and gamma-ray bands \citep[see][and references therein]{2015SSRv..191..315M}. Recent observational results and analysis suggest magnetar activity likely comprise at least some fraction of already-detected FXTs \citep[e.g.,][]{2025MNRAS.537..931D}  or whose activity may be detectable as FXTs in the local universe \citep[e.g.,][]{2025ApJ...989...63H,2025NatAs...9..111P}.  

Magnetars in our Galaxy typically reside in a quiescent state. From time to time they exhibit short-duration bursts (lasting milliseconds to seconds) and occasionally burst storms---a series of bursts occurring over minutes to days. These episodes often accompany a longer-lived outburst phase of higher persistent emission, which can last weeks to months \citep[see][for a recent review of magnetars' transient activity]{2024FrASS..1188953N}.

Separately, giant flares (MGFs) represent rare, extreme, but not cataclysmic events, characterized by a brief, intense spike (lasting milliseconds) followed by a modulated tail that decays over several minutes. MGFs are likely associated with stellar crust activity \citep[e.g.,][]{1984Natur.310..121L,1995MNRAS.275..255T,2015MNRAS.449.2047L,2016ApJ...824L..21L,2023ApJ...947L..16L, 2025arXiv250813419B}, possibly as the solid crust quakes caused by evolution and build-up of stresses by the strong magnetic field. Three nearby Galactic events have been observed (2 from the Milky Way, 1 from the LMC); while recently a total of six extragalactic MGF candidates have been identified \citep{burns2021identification, trigg2023grb, 2025A&A...694A.323T, roberts2021rapid,2024Natur.629...58M,2025ApJ...979L..25R}.
The spike reaches $E_{\rm iso}\sim10^{44}-10^{46}$ erg and its energy spectrum,  $E^2 dN/dE$, peaks in the soft gamma rays, evolving in time from 1 MeV to 0.3 MeV with a hard photon power-law index of $\sim$0). 

The initial bright, hard spike is typically followed by a significantly fainter, softer tail that can persist for several minutes, exhibiting flux variability modulated by the neutron star's spin period \citep[][]{2008AandARv..15..225M}. This pulsating tail spectrally peaks in the hard X-ray band, from a blackbody temperature of $kT \sim$ 10 keV to 20 keV, cooling in time to a few keV before shutting off \citep{Feroci_2001,Hurley+05}. The average luminosity is around $10^{41}–10^{42}$ erg/s indicating apparent blackbody areas similar to that of a neutron star. Thus, the tails might be approximate standard candles \citep[see table 2 of][]{2008AandARv..15..225M}. While these tails have been observed in Galactic MGFs, no definitive extragalactic counterpart has been detected so far, likely due to their faintness \citep[although, see][for a candidate]{2025MNRAS.537..931D}. Orphan MGF tails may arise when the collimated jet emission responsible for the initial spike is relativistically beamed away from the observer’s line of sight. To date, no orphan tails have been observed, likely due to the limited sensitivity of current instruments, which is insufficient to detect (let alone characterize) such events.

Beyond the pulsating tails, extended X-ray emission following MGFs has gained renewed interest as a potential site for r-process nucleosynthesis. The very recent study \citep{2025ApJ...984L..29P} suggests that this emission, lasting thousands of seconds, could be linked to the production of neutron-rich heavy elements such as gold, due to the process generating the impulsive spike. 

\begin{figure*}
    \centering
    \includegraphics[width=0.58\linewidth]{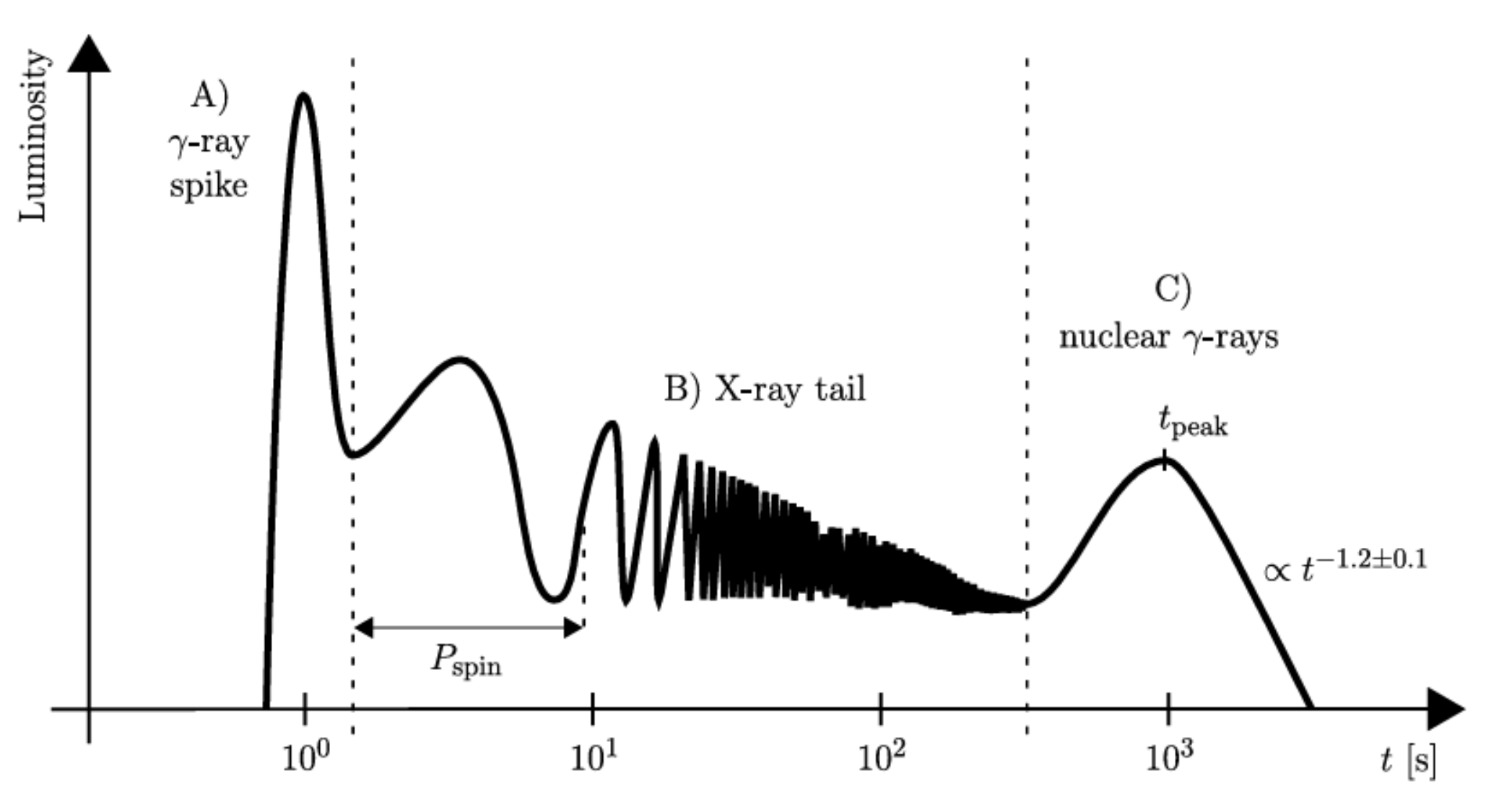}
    \caption{Schematic light curve of a magnetar giant flare. The emission proceeds in three phases: (A) an initial $\gamma$-ray spike of sub-second duration, (B) a pulsating X-ray tail modulated at the neutron star spin period, and (C) a delayed nuclear $\gamma$-ray component peaking at $t_{\rm peak}$ and decaying approximately as a power law of index $-1.2$. Credits: Jakub Cehula and Anirudh Patel.}
    \label{fig:mgf_scheme}
\end{figure*}

Figure \ref{fig:mgf_scheme} provides a schematic illustration of the three phases of an MGF emission that we will consider in this work: the bright, short spike, the modulating decaying tail, and the radioactively powered (r-process) emission at later times.

The Advanced X-ray Imaging Satellite (AXIS) \citep{2023SPIE12678E..1ER} is a recently selected NASA Probe-class mission currently in Phase A. AXIS is designed to provide an order-of-magnitude improvement in sensitivity over \textit{Chandra}, combined with a large field of view, rapid response, and excellent angular resolution. In this work we are interested in assessing AXIS capabilities in the context of FXTs, in particular, we consider all three phases of the MGF emission and assess observational prospects.

We focus on possible serendipitous detection of the short MGF spike during AXIS's prime mission period in Sec.~\ref{sec:spike}. 

In Sec.~\ref{sec:tail}, noting that, with a rapid response \textit{a\' la Swift}, the superior sensitivity of AXIS could provide the first opportunity to probe MGFs' periodic tail from local ($<30$~Mpc) extragalactic MGFs, we determine the optimal time-to-target (TtT) that would be required to detect and characterize such signals at extragalactic distances. 


In Sec.~\ref{sec:rproc}, we carry out a preliminary, simplified assessment aimed at exploring the prospects for detecting soft X-ray emission associated with an r-process signature. Specifically, we consider the decay of parent nuclei that emit lines in the AXIS energy band. This exercise is intended only as a quick feasibility study to determine whether pursuing a more detailed investigation would be scientifically worthwhile.

Summary and conclusions are provided in Sec.~\ref{sec:conclusion}.

\section{MGF serendipitous detection}
\label{sec:spike}
Assuming a signal-dominated burst lasting 64\,ms, we adopt the AXIS sensitivity for transient detection as 
$2.2 \times 10^{-10}\,\mathrm{erg\,cm^{-2}\,s^{-1}}$. 
This is obtained from the deep (7 Ms) field sensitivity in the 0.3--3\,keV 
($\sim 10^{-18}\,\mathrm{erg\,\,cm^{-2}\,s^{-1}}$, \citep{2023SPIE12678E..1ER}) rescaled to 64\,ms. 
We aim to estimate the number of MGFs AXIS could serendipitously detect within its FoV in the 5 years of prime mission, 
and the maximum distance at which they could be detected.

To do this, we assume the local ($z=0$) intrinsic volumetric rate as measured by \cite{burns2021identification}\, to be 
$3.8 \times 10^{5}\,\,\mathrm{Gpc^{-3}\,yr^{-1}}$ for a spike of $E_{\rm iso} > 3.7 \times 10^{44}\,\mathrm{erg}$.

Firstly, because the MGF spectral and energetic quantities are bolometric, we need to know the conversion factor to go from bolometric quantities to AXIS energy range. Assuming a typical observed Comptonized spectrum of MGFs (power-law index of 0.6, peak energy 290 keV) the ratio between the integrated flux in AXIS 0.3-10 keV band and the bolometric (1-10000 keV) band is $f_{\rm bol2AXIS} \sim 0.014$. The maximum bolometric energy observed so far is $E_{iso, bol}\sim5.75\times10^{46}$ erg \citep{burns2021identification}, corresponding to a maximum luminosity in the AXIS band (considering a duration of $\sim64$ ms) of
$L_{\rm iso,AXIS}^{\rm max} = L_{\rm iso,bol}^{\rm max} \times f_{\rm bol2AXIS} \sim 1.3\times10^{46}$ erg/s, where $L_{\rm iso,bol}^{\rm max}=E_{\rm iso, bol}/64$~ms. We note that the assumption that the spectrum in the soft X-ray band is the extrapolation of the gamma-ray Comptonized spectrum is an educated guess, but no direct observations are available. As pointed out in \cite{2013ApJ...775L..34R}, observations of ionospheric disturbances from the SGR~1900+14 giant flare suggest that the soft X-ray flux (3–10 keV) can exceed the hard X-ray level by nearly an order of magnitude \citep{1999GeoRL..26.3357I}, thus our assumption can be considered conservative.

Secondly, considering an effective observation factor of 95\% (i.e., removing the time spent slewing/calibrating) and accounting for a 450 arcmin$^2$ ($3.8 \times 10^{-5}$~sr) field of view, we define an AXIS total observing efficiency as $\epsilon_{\rm AXIS} = 0.95 \Omega_{\rm FoV}$, where $\Omega_{\rm FoV} = \frac{\rm FoV [sr]}{4\pi}$.

Now we can define the differential rate of AXIS observations of a MGF at a given distance as
\begin{equation}
\frac{dN^{\mathrm{yr}}_{\mathrm{obs}}}{dD_L}(D_L) = \mathcal{R}_{\mathrm{vol}} \times \frac{dV_c}{dD_L}(D_L) \times \epsilon_{\rm AXIS}
\end{equation}
where $\frac{dV_c}{dD_L}(D_L)$ is the comoving volume in the $dD_L$-thin spherical shell at the distance $D_L$, and $0<D_L< D_{L}^{\rm max}$ is the luminosity distance in AXIS energy range, and 
\begin{equation}
    D_{\rm L}^{\mathrm max} = \sqrt{\frac{L^{\mathrm max}_{\mathrm iso}}{4\pi \phi^{\mathrm AXIS}_{\mathrm sens}}}
\end{equation}
obtaining $D^{\rm max}_{\rm L,AXIS} \sim 692 {\rm~Mpc}$.

\begin{figure*}
    \centering
    \includegraphics[height=0.25\linewidth]{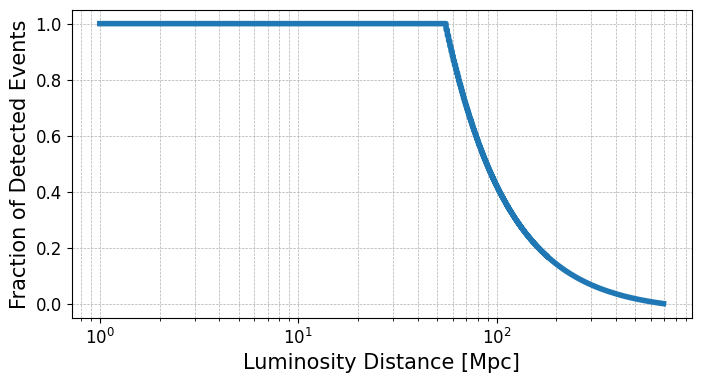}~~~~
    \includegraphics[height=0.25\linewidth]{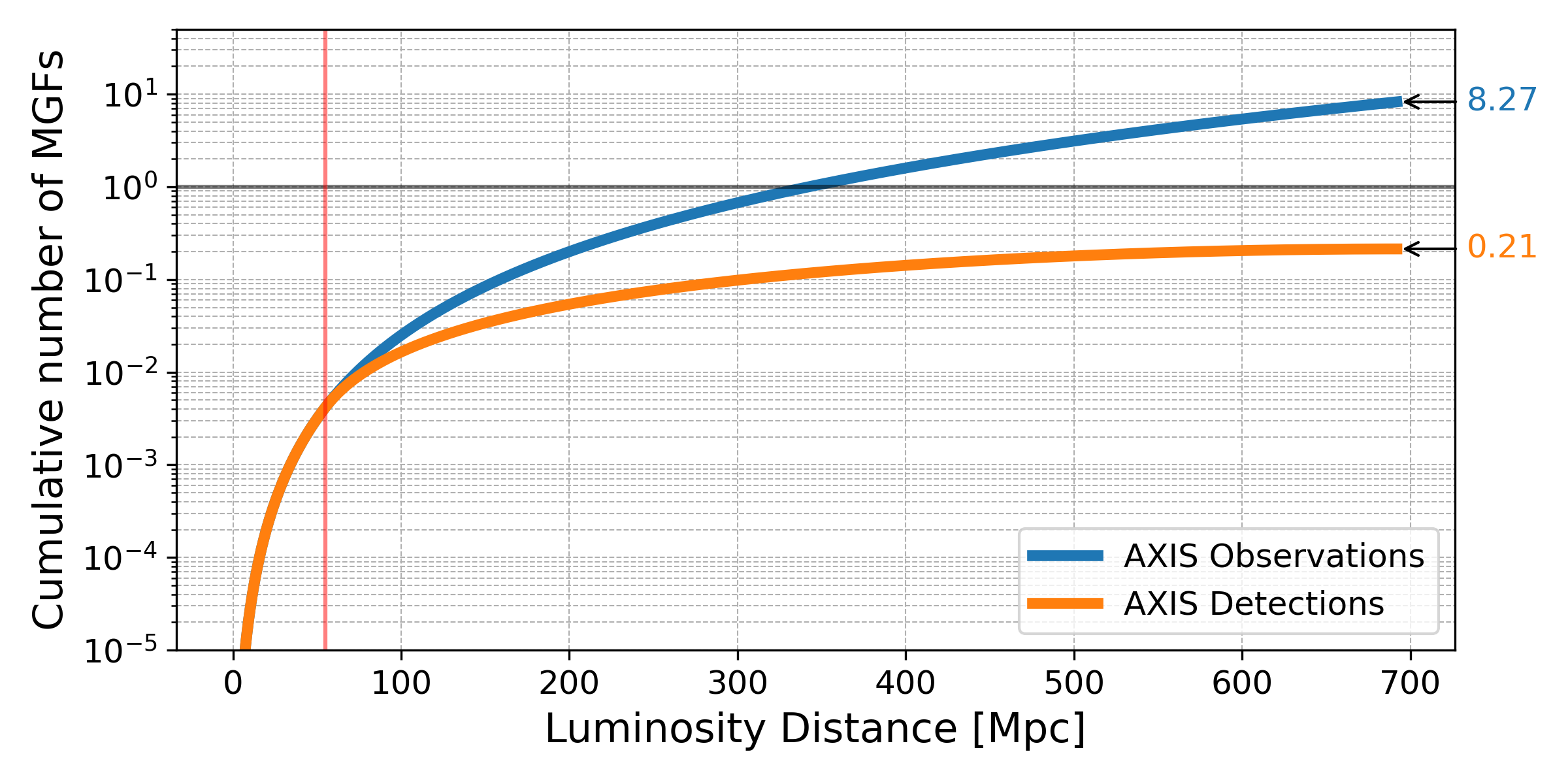}
    \caption{\textit{Left:} Fraction of AXIS detectable events as function of the distance. \textit{Right}: Cumulative number of MGFs as seen by AXIS in 5 years mission as a function of the distance.}
    \label{fig:frac_detections}
\end{figure*}

We now need to calculate the fraction of events detected (i.e., those with flux above the AXIS sensitivity) at a given distance, $f_{\mathrm{det}}(D_L)$, accounting for the intrinsic energy distribution, which follows a power law with index $-1.7$ \citep{burns2021identification}. While intrinsically faint events are far more numerous than bright ones, they quickly fall below the detection threshold as distance increases. Fig.~\ref{fig:frac_detections} shows the fraction of detected events as function of the distance.

Now we can compute the total number of detected events per year as 
$$\frac{dN^{\mathrm{yr}}_{\mathrm{det}}}{dD_L}(D_L) = \frac{dN^{\mathrm{yr}}_{\mathrm{obs}}}{dD_L}(D_L) \times f_{\mathrm{det}}(D_L)$$

Multiplying by 5, corresponding to the 5-year prime mission duration, we can obtain the cumulative number of events out to the maximum distance AXIS would detect. Fig.~\ref{fig:frac_detections} (right) shows the cumulative number of MGF events over 5 years that will fall within AXIS’s field of view (blue curve) and those that would be detected, being above the AXIS sensitivity threshold (orange curve). AXIS detection prediction in 5-years of mission is $N^{\rm MGFs}_{\rm AXIS \, \, 5 yr} = 0.21^{+0.44}_{-0.04}$.

Note, this calculation ignores the greater hardness for brighter events. We assumed an average Comptonized spectrum (index = 0.6, energy peak = 250 keV), but a harder spectrum (i.e., index=0.0, energy peak = 1500 keV) would result in a slightly lower $f_{bol2AXIS}$. In the worst-case scenario that all MGFs have harder spectra (which we know is not the case) we get only a minor change in the final estimate: $N^{\rm MGFs}_{\rm AXIS \,\, 5 yr} = 0.19^{+0.39}_{-0.04}$.  Also ignored in this calculation is the possible evolution of the rate with redshift. Such evolution is not known exactly, but assuming that MGFs trace massive star evolution, we estimate that accounting for this effect would only improve the serendipitous detection rate by a $\sim10\%$.

In conclusion, AXIS could detect all MGFs out to 55 Mpc, the brightest 40\% out to 100 Mpc and the brightest 10\% within 300 Mpc. Unfortunately, given the relatively local distances  it is quite unlikely that AXIS could serendipitously observe such fast transients in its FoV in 5 years. 

It is worth noting that AXIS would gain a significant advantage by devoting a fraction of its observing time to nearby galaxies ($<$100 Mpc), particularly star-forming systems. In this regime, the volumetric average assumption breaks down, since a small number of galaxies with high star-formation activity can dominate the overall event rate \citep{2025ApJ...980..211B}. Targeted monitoring of such galaxies would therefore enhance the chances of capturing MGFs and, more generally, fast X-ray transients, serendipitously above the isotropic expectation.
Within the Local Group, the Milky Way, LMC, SMC, M31, and M33 together contribute a combined star formation rate of about 3.5 $M_\odot$ yr$^{-1}$. If gamma-ray detections and identifications are complete out to these distances---which is plausible, provided timely follow-up and robust identification strategies---the corresponding MGF rate would be approximately 0.2 events per year. This translates into the possibility of detecting roughly one MGF over a five-year mission: an exciting prospect, though still uncertain.

\subsection{Generic framework to assess FXT detectability}

\begin{table*}[ht]
\centering
\begin{tabular}{|p{6cm}|p{6cm}|p{5cm}|}
\hline
\textbf{Instrument information needed} & \textbf{Transient information needed} & \textbf{Assumptions} \\
\hline
Sensitivity for transient detection (flux threshold) & Burst duration & Burst duration is fixed and representative of the transient class \\
\hline
Energy band of the instrument & Intrinsic volumetric rate (local or redshift-dependent) & Volumetric rate is assumed constant (unless specified otherwise) \\
\hline
Field of view & Maximum isotropic equivalent energy (or luminosity) observed & Spectrum is assumed representative of the whole population \\
\hline
Effective observing factor (e.g., Sun/anti-Sun obscuration, calibration/slewing losses) & Spectral shape (e.g., power-law, Comptonized) and typical parameters (index, peak energy) & Spectral parameters do not strongly evolve with brightness \\
\hline
Observing efficiency: combination of duty cycle and FoV coverage & Luminosity/energy distribution (e.g., power-law index) & No cosmological evolution of event rate (unless modeled explicitly) \\
\hline
Maximum detectable distance derived from sensitivity and luminosity & Fraction of events detectable above threshold as function of distance & Harder/softer spectra only modestly affect detectability \\
\hline
 & Expected detection rate over the mission lifetime & Population properties (spectrum, duration, rate) are stationary in time \\
\hline
\end{tabular}
\caption{General framework for estimating the detectability of transient events by an instrument. }
\label{tab:transient_rate}
\end{table*}

The exercise made above for the case of MGFs could be applied to other type of short transients, where the term `short' can be interpreted as broadly as any transient with duration such that the assumption of negligible background still holds. AXIS's background rate is estimated to be around $4.43\times10^{-8}$ counts/s, which can be considered negligible ($<10^{-3}$ counts) up to $10^5$ s exposure. Because this could be relevant to other groups, possibly beyond the specifics of AXIS we provide a summary of needed quantities in Tab.~\ref{tab:transient_rate} and assumptions needed if one would like to repeat this calculation for different events and missions. The generic workflow looks like this:

\begin{enumerate}
    \item \textbf{Instrument characterization:}  
    Determine the sensitivity (flux threshold), energy band, field of view, and effective observing factor (e.g., duty cycle, slewing losses).  
    Combine these into an overall observing efficiency.
    
    \item \textbf{Transient population properties:}  
    Specify the burst duration, intrinsic volumetric rate (local or redshift-dependent), maximum observed isotropic energy (or luminosity), and typical spectral shape with characteristic parameters (e.g., index, peak energy).  
    Define the luminosity or energy distribution (e.g., power-law).
    
    \item \textbf{Detection horizon:}  
    Compute the maximum detectable distance by comparing transient luminosity to the instrument sensitivity in the relevant energy band.
    
    \item \textbf{Detection fraction:}  
    For each distance, estimate the fraction of events above the sensitivity threshold, accounting for the luminosity/energy distribution.
    
    \item \textbf{Event rate calculation:}  
    Multiply the comoving volume element, volumetric rate, and instrument efficiency to obtain the differential observed rate.  
    Apply the detection fraction to obtain the differential detection rate.
    
    \item \textbf{Cumulative detections:}  
    Integrate over distance (or redshift) and multiply by the mission lifetime to estimate the total number of expected detections.
    
    \item \textbf{Assumptions check:}  
    Clearly state the assumptions on burst duration, spectral shape, population evolution, and efficiency factors.  
    Discuss possible systematic uncertainties (e.g., harder/softer spectra, redshift evolution of the rate).
\end{enumerate}

\section{MGF tails detection}
\label{sec:tail}

Ref.~\citet{2023arXiv231107673S} estimated that, based on its sensitivity, AXIS could detect magnetar candidates out to a distance of $\sim$3.5 Mpc in a deep 250 ks exposure, via persistent emission. However, confirming such detections through periodicity searches would be extremely challenging in the absence of \textit{Athena} \citep{2025NatAs...9...36C}. At present, the most promising strategy appears to be the detection of the tail emission of a giant flare or an `intermediate flare' outburst \citep{2025ApJ...989...63H}. The arcsecond-scale localization provided by AXIS would further enable studies of the environments of extragalactic magnetars, likely associated with star-forming regions, as observed in the Milky Way, but potentially with unexpected variations.

Given the high scientific impact resulting from being able to study extragalactic magnetars, we investigate the feasibility of repointing AXIS in response to externally triggered MGFs detected by gamma-ray burst monitors. In particular, we assess how rapidly the observatory would need to slew in order to enable both a robust detection\footnote{That is, obtaining a signal clearly distinguishable from the background with a meaningful characterization.} by collecting sufficient photons to identify the pulsations.

\begin{figure*}
    \centering
    \includegraphics[height=0.25\textwidth]{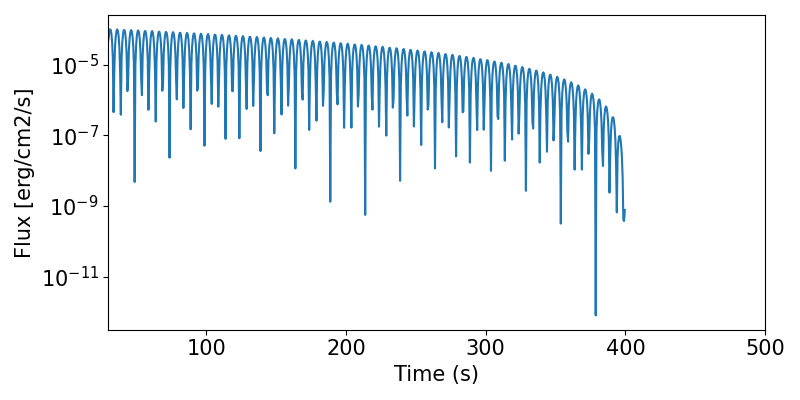}~~~~
    \includegraphics[height=0.25\textwidth]{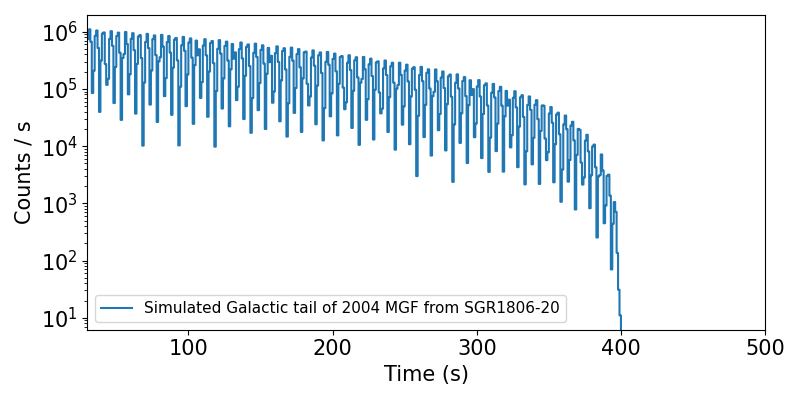}
    \caption{\textit{Left:} Lightcurve of an MGF tail mimicking the one observed during the 2004 event from SGR 1806–20 \citep{Hurley+05}. \textit{Right:} Simulated AXIS light curve of a decaying tail similar to the 2004 event from SGR 1806–20.}
    \label{fig:mgf_2}
\end{figure*}

To determine the detectability of extragalactic MGF tails by AXIS we take the observed tail of the 2004 giant flare from SGR 1806–20 \citep{Hurley+05} and simulate an AXIS observation assuming the same spectrum and the one measured for the event and accounting for the AXIS energy dependent effective area.

The simulation assumes a modulated blackbody tail with a $5$~second period, a pulse fraction of 100\%\footnote{In the literature \citep{Hurley+05,2007AstL...33....1F} the tail is described as ``highly modulated'', but no quantitative estimate of the pulse fraction is reported.}, and an overall decay consistent with a trapped fireball scenario:
\begin{equation}
L_X(t) = L_0~\left(1 - \frac{t}{t_{\rm evap}}\right) \frac{a}{1-a}
\end{equation}
where $L_0$ is the initial luminosity of the tail, $t_{evap}$ is the evaporation timescale (set to 400) and $a$ is a constant (set to 0.606).

The flux of the first peak was converted to AXIS count rate using {\tt webPIMMS}\footnote{\url{https://heasarc.gsfc.nasa.gov/cgi-bin/Tools/w3pimms/w3pimms.pl}} assuming an unabsorbed flux of $1.31\times10^{-4}$ erg cm$^{-2}$ s$^{-1}$ ($20-100$ keV) as input\footnote{The flux $1.31 \times10^{-4}$ erg cm$^{-2}$ s$^{-1}$ (20–100 keV) is derived from the intrinsic isotropic luminosity estimated at the beginning of the tail by \cite{Hurley+05} and assuming a distance of the magnetar of 8 kpc. We note that this distance is debated in literature. We adopted the value 8 kpc as estimated  in \cite{2008MNRAS.386L..23B}, which is nearer than other values found in previous literature. Because assuming a closer source results in a lower intrinsic luminosity, our choice can be considered conservative.}, a blackbody temperature of $kT=8$ keV, and a Galactic absorption column of $N_H = 10^{20}$ cm$^{-2}$, according to literature values. {\tt webPIMMS} estimates an initial count rate of $1.1\times10^6$ counts/s on-axis (and $\sim8.3 \times 10^5$ counts/s averaged over the AXIS field of view) in the 0.3–10 keV band.

\begin{figure*}
    \centering
    \includegraphics[height=0.25\textwidth]{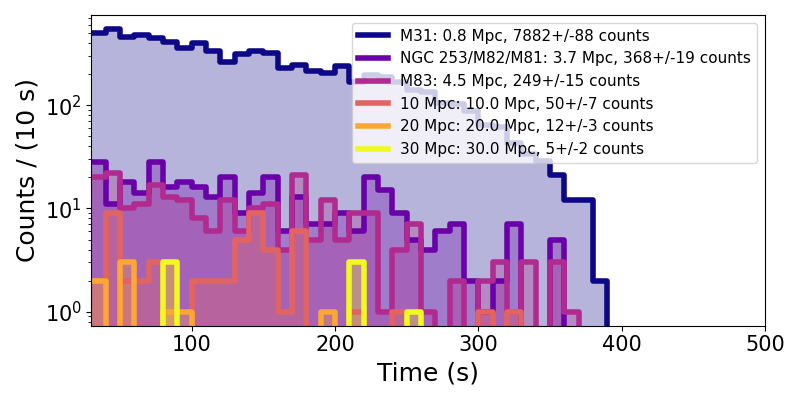}~~~~
    \includegraphics[height=0.25\textwidth]{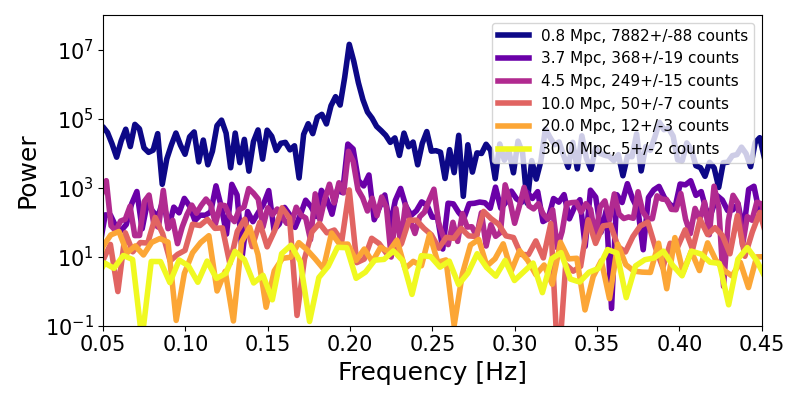}
    \caption{\textit{Left:} Simulated AXIS-detected tails of a giant flare similar to the 2004 SGR 1806–20 event, scaled to different extragalactic distances. \textit{Right} Power spectra of simulated AXIS light curves for a magnetar giant flare tail observed at increasing distances. The evident peak $f=0.20$ corresponds to the 5 s periodicity that we input in the simulation.}
    \label{fig:mgf_3}
\end{figure*}

The simulated AXIS-detected tail is then scaled to increasing extragalactic distances from 0.8 Mpc (M31) to 30 Mpc. For each case, Poisson fluctuations were added to the binned light curve. The resulting expected AXIS counts (per 10 s binning) are shown in Fig.\ref{fig:mgf_3} (left) for representative galaxies: M31 (0.8 Mpc), NGC 253/M82/M81 ($\sim3.7$ Mpc), M83 (4.5 Mpc), as well as for generic sources placed at 10, 20, and 30 Mpc. Also in this case the duration of the tail is short enough to allow us to assume a negligible background rate: $\sim4 \times 10^{-8}$ counts/s in a 0.6"-radius circular region around the transient location. The background was estimated from the AXIS simulated full-FoV background spectrum, integrated between 0.3–10 keV and rescaled for the selected region). With a fast repointing similar to the Neil Gehrels Swift Observatory \citep{2004ApJ...611.1005G}, tail detections could be possible up to $\sim20$ Mpc.
The power spectra are computed from the simulated AXIS event lists and adopting a 0.2 second binning, assuming AXIS time resolution to be $\sim$5 frames/s \citep{2023SPIE12678E..1ER}. Fig.~\ref{fig:mgf_3} (right) shows the power spectra for a periodic decaying tail from a MGF at different distances. The input signal at 5-second periodic modulation is evident, yielding a peak in the power spectra near 0.2 Hz. As distance increases and photon statistics degrade, the periodic signal becomes progressively less detectable. While the periodicity is prominent in nearby galaxies (e.g., M31 at 0.8 Mpc), it becomes indistinguishable from noise beyond $\sim15-20$ Mpc.

\begin{figure*}
    \centering
    \includegraphics[height=0.24\textwidth]{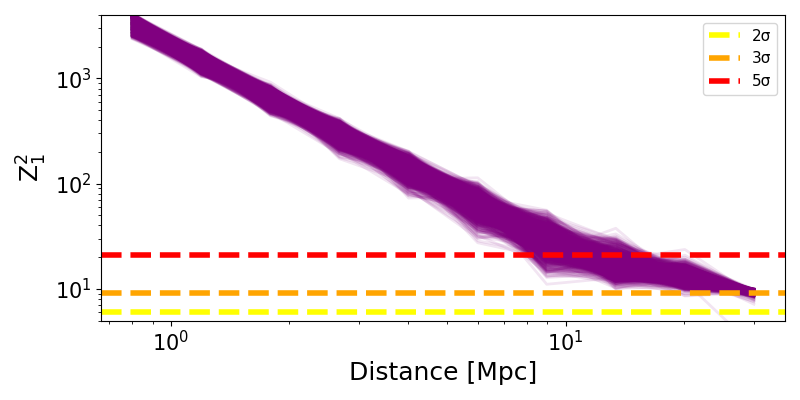}
    \includegraphics[height=0.24\textwidth]{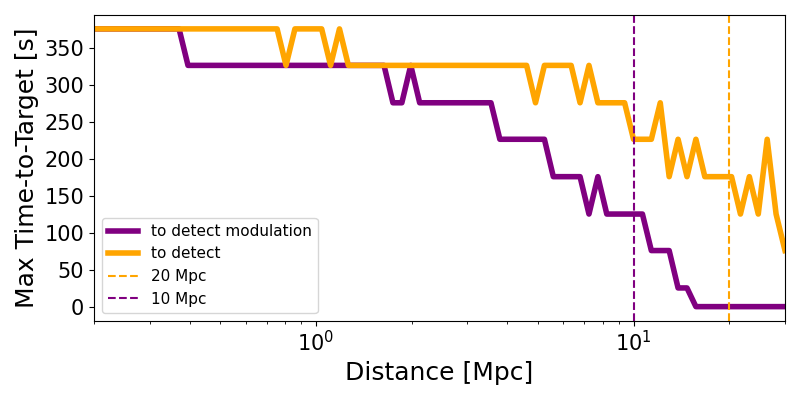}
    \caption{\textit{Left:} Periodicity detection significance as a function of distance for simulated AXIS observations of magnetar giant flare tails. \textit{Right:} Maximum time-to-target on a magnetar giant flare as a function of distance, based on simulated AXIS observations.}
    \label{fig:mgf_4}
\end{figure*}

We ran 1000 simulations for each distance where each curve represents a Monte Carlo realization of the modulated tail obtained as described above and allowing for Poissonian fluctuations in the number of the counts in each time bin. We analyzed each power spectrum using the $Z^2_1$ test for periodicity. The purple band in Fig.~\ref{fig:mgf_4} therefore shows the spread in $Z^2_1$ values due to Poisson fluctuations. The horizontal dashed lines indicate the thresholds for $2\sigma$, $3\sigma$, and $5\sigma$ detection significance. Periodicity remains detectable above the $5\sigma$ level out to distances of $\sim8–10$ Mpc under the assumed conditions. The detection of additional quasi-periodic oscillations, which have been claimed in past Galactic MGFs tails \citep[see, e.g.,][]{2006MNRAS.368L..35L, watts2007neutron,2009MNRAS.396.1441C} and associated with neutron star crustal modes, would require a time resolution $\lesssim$ms. 

To consider is the time delay from the beginning of an MGF tail emission and the instant AXIS would be on target and start observing. We call this delay the time-to-target (TtT). In particular we are interested in the maximum TtT that would allows a detection of the typical seconds-long modulation of MGF tails as a function of the distance. Fig.~\ref{fig:mgf_4} (right), shows exactly this. The orange curve indicates the maximum time within which a transient can be detected to confidently identify it as an MGF (minimum of $\sim$2 counts detected with zero background expected). The purple curve shows the more stringent condition required to detect periodic modulation in the tail at about 3 sigma (minimum of $\sim$40 counts detected). Vertical dashed lines mark reference distances of 10 Mpc and 20 Mpc. The step-like features and fluctuations in both curves arise from noise in the simulations. 

AXIS can identify an MGF tail out to $\sim20$ Mpc detecting at least 2 counts, with a $TtT \sim 200$ s, and can detect ($3\sigma$) modulation of MGF tail out to $\sim5$ Mpc detecting at least 40 cts, with a $TtT\sim150$ s. Such fast turnaround is currently possible with the technology onboard the Neil Gehrels Swift Observatory.

Once again, we want to remark how adopting an observing strategy in which nearby galaxies ($<$50 Mpc) fall in AXIS's FoV, could boost the chance of detecting orphan tails serendipitously, offering crucial new observations to shed light on the physics, collimation/beaming, and geometry of these events.

\section{R-processes from MGFs}
\label{sec:rproc}

MGFs have also been suggested as potential sites for rapid neutron capture process (r-process) nucleosynthesis, responsible for forming many of the heaviest elements in the Universe \citep{2024MNRAS.528.5323C}. The synchrotron radio afterglow following the prompt gamma-ray flare indicates an associated sub-relativistic baryon-loaded ejecta component (e.g., \citealt{Gelfand+05,Granot+06}). Originating from the magnetar itself, the ejecta are sufficiently neutron-rich such that neutron captures proceed faster than beta-decays, enabling transmutation of light nuclei into heavier trans-iron species \citep{Patel+2025b}.  Recently \cite{2025ApJ...984L..29P} demonstrated direct evidence for this process, utilizing the archival data taken during the 2004 MGF from SGR 1806-20 with the anticoincidence shield on the INTEGRAL satellite \citep{2005ApJ...624L.105M}, the rear detectors on RHESSI \citep{Boggs2007}, and the Konus gamma-ray spectrometer on Wind \citep{frederiks2007giant}.

\begin{figure*}
    \centering
    \includegraphics[width=0.4\linewidth]{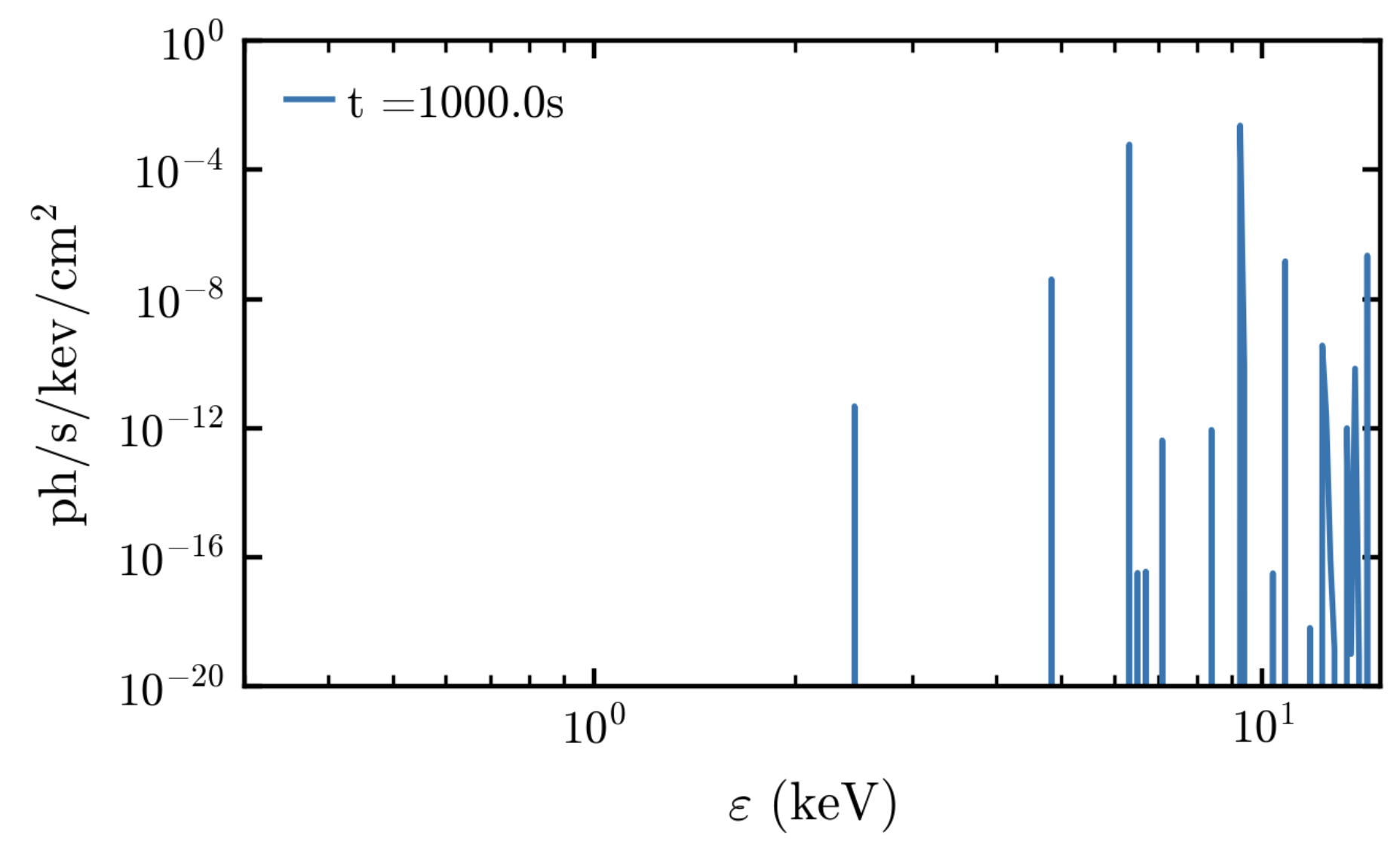}~~~
    \includegraphics[width=0.4\linewidth]{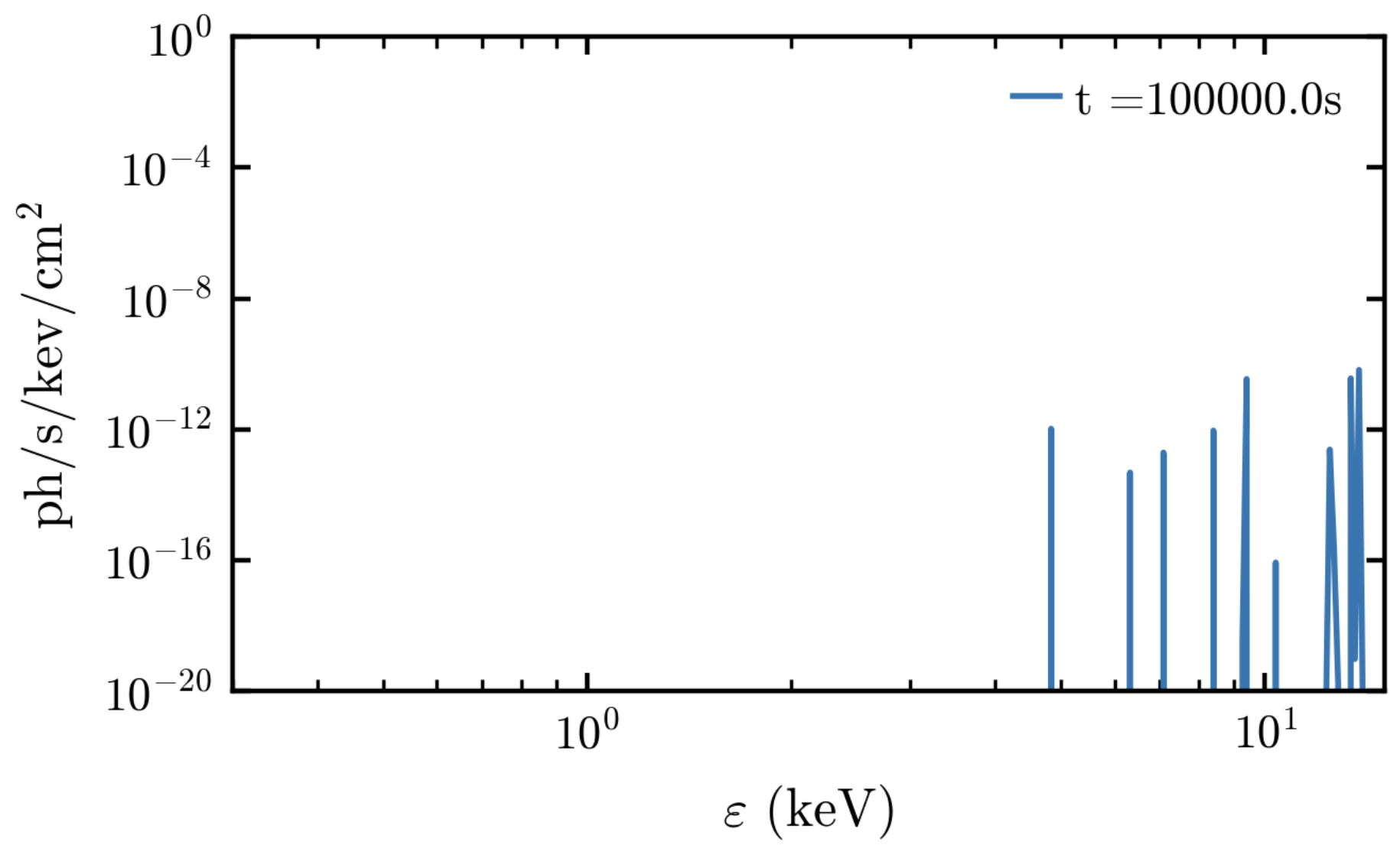}
    \includegraphics[width=0.4\linewidth]{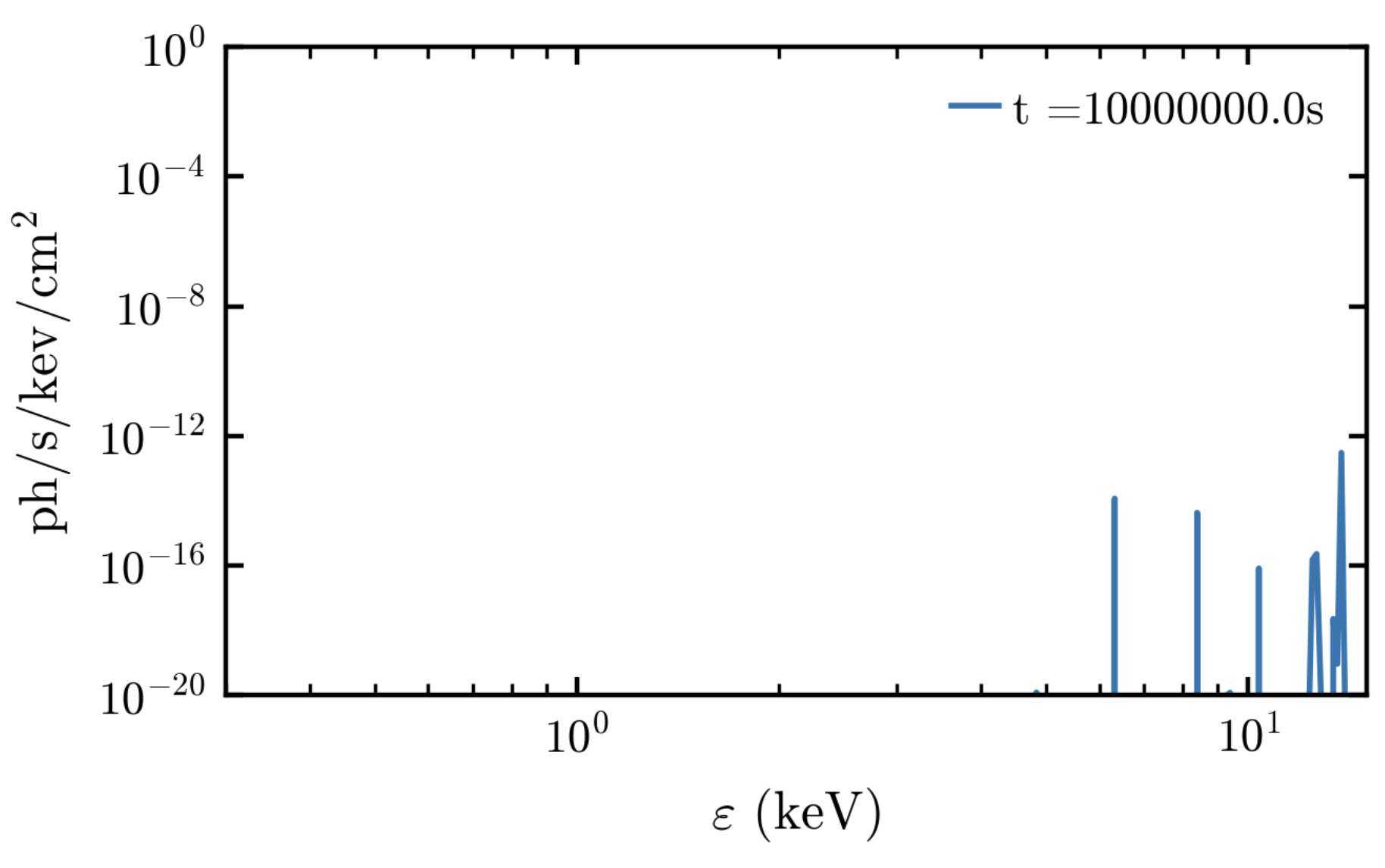}~~~
    \includegraphics[width=0.4\linewidth]{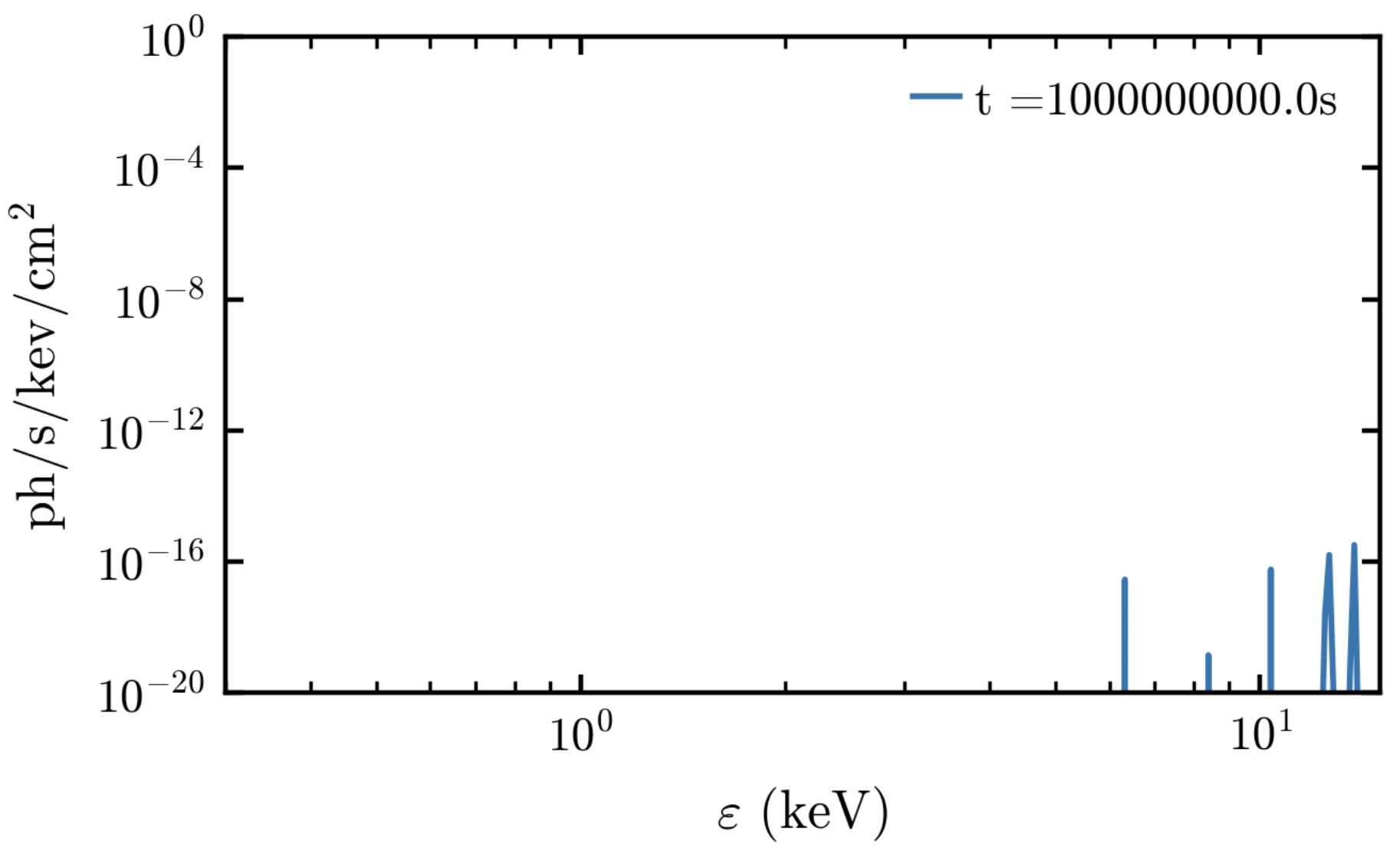}
    \caption{X-ray predictions below at t = $10^3$ s, $10^5$ s, $10^7$ s (4 mo), and $10^9$ (32 yrs) for a source at 10 kpc. Equivalent to Figure 3 of \cite{2025ApJ...984L..29P} but for different timescales after the event.}
    \label{fig:mgf-5}
\end{figure*}

Although the r-process decay spectrum peaks at $\sim$MeV energies, several decay lines occur at X-ray energies $\sim 10$ keV \citep{2014MNRAS.438.3243R}, albeit carrying a luminosity roughly 2--3 orders of magnitude smaller than the total. The four panels in Fig.~\ref{fig:mgf-5} show the soft-X-ray r-process lines expected in MGFs, as would appear after different times from the main event (bright, short peak). Two key factors make X-ray detection significantly more challenging than in the MeV gamma-ray regime. First, only a small fraction of the radioactive decay energy—on the order of $\sim1$\%—is emitted in the X-ray band, compared to the gamma-ray output (see Fig. 3 of \citealt{2025ApJ...984L..29P}). Second, the high photoelectric absorption opacity of the ejecta severely attenuates X-rays (Barnes et al. 2016), delaying their escape until roughly $3\times10^4$ s post-explosion—much later than the $\sim10^3$ s transparency timescale for gamma rays. By that time, the radioactive power has already declined substantially. While a more quantitative treatment would require detailed Monte Carlo modeling, current estimates suggest the X-ray luminosity will peak at $\sim10^{34} - 10^{35}$ erg/s, making such signals extremely difficult to detect beyond the Milky Way. In order-of-magnitude terms, assuming a soft X-ray luminosity for r-process in MGFs on the bright side of $\sim10^{35}$ erg/s, at the distance of M31 (Andromeda) the expected flux would be approximately $2\times10^{-15}$ erg cm$^{-2}$ s$^{-1}$. We consider how these values compares to the expected background. As mentioned before, a background rate of approximately $4\times 10^{-8}$ counts/s within a $0.6^{\prime\prime}$-radius circular region centered on a transient is effectively negligible, even over a 10 ks observation. The AXIS sensitivity can be then easily rescaled, obtaining $\approx1.4 \times 10^{-15}$ erg cm$^{-2}$ s$^{-1}$ for an exposure of 10 ks. If temporal binning is needed—for example, using three time bins to characterize a temporal decay—the effective sensitivity per bin degrades to $\approx4.2\times10^{-15}$ erg cm$^{-2}$ s$^{-1}$. This number is close to the expected signal, which makes it hard, given the approximations involved, to establish with certainty if such a measurement would be feasible.

We can consider the single line emission. Firstly, we look at the case of 32 years after the event (bottom-right panel of Fig.~\ref{fig:mgf-5}) to see if AXIS, once launched could attempt an observation of the late r-process emission from SGR 1806-20, we focus on the $\sim6.1$ keV line with a flux of $\sim36\times10^{-16}$ ph/cm$^2$/s, corresponding to an energy flux of $\sim 5.7 \times 10^{-24}$ erg/cm$^2$/s. Inserting this flux into {\tt webPIMMS} we predict an AXIS photon rate of $3.56 \times 10^{-13}$ counts/s with AXIS FoV-average in the 6--6.2 keV range. This number is $\sim5$ orders of magnitude lower than the estimated background, suggesting that AXIS would not be able to detect such late-time emission.

Secondly, we can consider the case of a Galactic event. Such events are rare (approximately one every 10--30 years, based on the observed rate). However, since the last event was detected in 2004, it would not be unreasonable to expect another in the near future. Referring to the top-right panel of Fig.~\ref{fig:mgf-5}, and focusing on the feature around 5 keV, a 10 ks exposure yields a webPIMMS-predicted count rate of $6.1 \times 10^{-7}$ cps in the 4.8–5.0 keV band. This flux level would be readily detectable by AXIS; however, opacity could still be significant at this stage, potentially blocking X-rays from escaping.

Note, in the modeling reported in Fig.~\ref{fig:mgf-5} there is no Doppler broadening or radiative transport included, which will affect the early time spectra. Late time spectra are more accurate in this regard, however the predicted flux is low at that point in time.

\section{Conclusions}
\label{sec:conclusion}
Magnetar giant flares provide a rare but powerful probe of extreme physics, including magnetic-field dissipation and triggering mechanisms, synthesis of heavy elements via r-process nucleosynthesis, and quantum electrodynamic processes \citep{2025arXiv250203577B}. In this study, we assessed the detectability of MGFs with AXIS across multiple phases of their emission.

Serendipitous detections of short spikes are improbable within the AXIS field of view over the prime mission lifetime given their short duration and the assumption that the hard gamma-ray spectrum extends into the soft X-ray band, though bright events could in principle be observed out to several hundred Mpc, and the soft X-ray flux could be enhanced by a thermal component from an expanding fireball, as observed in short bursts and magnetar giant flare spikes.  
More broadly, we outlined a general workflow for estimating transient detectability with any instrument, combining sensitivity, volumetric rates, spectral assumptions, and mission efficiency. 

Pulsating X-ray tails represent a more promising avenue. Simulations demonstrate that AXIS could detect and characterize periodic modulations up to $\sim$5–10 Mpc, and identify tails as faint transients out to $\sim$20 Mpc, provided a rapid time-to-target response. Such detections would enable the first extragalactic measurements of magnetar periodic signals, offering new insight into fireball physics, emission geometry, and potential quasi-periodic oscillations (which would require ms-scale time resolution).

An optimal observing strategy for AXIS would be combining wide-field monitoring with targeted observations of nearby, star-forming galaxies ($<$100 Mpc). Such an approach would maximize the likelihood of detecting serendipitously the bright, short-lived spikes of MGFs, their fainter tails and potentially orphan tails. By concentrating on galaxies expected to dominate the local transient rate, AXIS could boost its scientific return beyond that achievable by uniform, random coverage.

X-ray emission related to r-process is predicted to be several orders of magnitude fainter than AXIS sensitivity at extragalactic distances. While not accessible beyond the Milky Way, targeted Galactic observations could place meaningful constraints on the role of MGFs in heavy-element production.

\section*{ACKNOWLEDGEMENTS}
We wish to thank the anonymous referee whose feedback helped improve the manuscript.
We also wish to acknowledge the AXIS team for providing the instrument response functions and sensitivity estimates, crucial to carry out this study.  The material is based upon work supported by NASA under award numbers 80GSFC21M0002 and 80GSFC24M0006. This work has made use of the NASA Astrophysics Data System. 

\bibliography{sample701}{}

\begin{thebibliography}{}
\expandafter\ifx\csname natexlab\endcsname\relax\def\natexlab#1{#1}\fi
\providecommand{\url}[1]{\href{#1}{#1}}
\providecommand{\dodoi}[1]{doi:~\href{http://doi.org/#1}{\nolinkurl{#1}}}
\providecommand{\doeprint}[1]{\href{http://ascl.net/#1}{\nolinkurl{http://ascl.net/#1}}}
\providecommand{\doarXiv}[1]{\href{https://arxiv.org/abs/#1}{\nolinkurl{https://arxiv.org/abs/#1}}}

\bibitem[{{Bauer} {et~al.}(2017){Bauer}, {Treister}, {Schawinski}, {Schulze}, {Luo}, {Alexander}, {Brandt}, {Comastri}, {Forster}, {Gilli}, {Kann}, {Maeda}, {Nomoto}, {Paolillo}, {Ranalli}, {Schneider}, {Shemmer}, {Tanaka}, {Tolstov}, {Tominaga}, {Tozzi}, {Vignali}, {Wang}, {Xue}, \& {Yang}}]{2017MNRAS.467.4841B}
{Bauer}, F.~E., {Treister}, E., {Schawinski}, K., {et~al.} 2017, \mnras, 467, 4841, \dodoi{10.1093/mnras/stx417}

\bibitem[{{Beniamini} {et~al.}(2025){Beniamini}, {Wadiasingh}, {Trigg}, {Chirenti}, {Burns}, {Younes}, {Negro}, \& {Granot}}]{2025ApJ...980..211B}
{Beniamini}, P., {Wadiasingh}, Z., {Trigg}, A., {et~al.} 2025, \apj, 980, 211, \dodoi{10.3847/1538-4357/ada947}

\bibitem[{{Bibby} {et~al.}(2008){Bibby}, {Crowther}, {Furness}, \& {Clark}}]{2008MNRAS.386L..23B}
{Bibby}, J.~L., {Crowther}, P.~A., {Furness}, J.~P., \& {Clark}, J.~S. 2008, \mnras, 386, L23, \dodoi{10.1111/j.1745-3933.2008.00453.x}

\bibitem[{{Boggs} {et~al.}(2007){Boggs}, {Zoglauer}, {Bellm}, {Hurley}, {Lin}, {Smith}, {Wigger}, \& {Hajdas}}]{Boggs2007}
{Boggs}, S.~E., {Zoglauer}, A., {Bellm}, E., {et~al.} 2007, ApJ, 661, 458, \dodoi{10.1086/516732}

\bibitem[{{Bransgrove} {et~al.}(2025){Bransgrove}, {Beloborodov}, \& {Levin}}]{2025arXiv250813419B}
{Bransgrove}, A., {Beloborodov}, A.~M., \& {Levin}, Y. 2025, arXiv e-prints, arXiv:2508.13419, \dodoi{10.48550/arXiv.2508.13419}

\bibitem[{{Burns} {et~al.}(2021){Burns}, {Svinkin}, {Hurley}, {Wadiasingh}, {Negro}, {Younes}, {Hamburg}, {Ridnaia}, {Cook}, {Cenko}, {Aloisi}, {Ashton}, {Baring}, {Briggs}, {Christensen}, {Frederiks}, {Goldstein}, {Hui}, {Kaplan}, {Kasliwal}, {Kocevski}, {Roberts}, {Savchenko}, {Tohuvavohu}, {Veres}, \& {Wilson-Hodge}}]{burns2021identification}
{Burns}, E., {Svinkin}, D., {Hurley}, K., {et~al.} 2021, ApJ Letters, 907, L28, \dodoi{10.3847/2041-8213/abd8c8}

\bibitem[{{Burns} {et~al.}(2025){Burns}, {Fryer}, {Agullo}, {Andrews}, {Aydi}, {Baring}, {Baron}, {Boorman}, {Boroumand}, {Borowski}, {Broekgaarden}, {Chandra}, {Chatzopoulos}, {Chen}, {Chipps}, {Civano}, {Comisso}, {C{\'a}rdenas-Avenda{\~n}o}, {Dang}, {Deibel}, {Eftekhari}, {Elliott}, {Foley}, {Fontes}, {Gall}, {Galleher}, {Gonzalez}, {Guo}, {Babiuc Hamilton}, {Harding}, {Henning}, {Herwig}, {Hix}, {Ho}, {Holley-Bockelmann}, {Hounsell}, {Hui}, {Humensky}, {Hungerford}, {Hynes}, {Jin}, {Johns}, {Gatu Johnson}, {Kennea}, {Kuranz}, {Lamb}, {Launey}, {Lewis}, {Liodakis}, {Livescu}, {Loch}, {MacDonald}, {Maccarone}, {Marcotulli}, {Meli}, {Messer}, {Miller}, {Milton}, {Most}, {Mumma}, {Mumpower}, {Negro}, {Neights}, {Nugent}, {Pasham}, {Radice}, {Rani}, {Read}, {Reifarth}, {Reily}, {Rhodes}, {Richard}, {Ricker}, {Roberts}, {Schatz}, {Shawhan}, {Takacs}, {Tomsick}, {Trigg}, {Urbatsch}, {Vassh}, {Villar}, {Wadiasingh}, {Waratkar}, \& {Zingale}}]{2025arXiv250203577B}
{Burns}, E., {Fryer}, C.~L., {Agullo}, I., {et~al.} 2025, arXiv e-prints, arXiv:2502.03577, \dodoi{10.48550/arXiv.2502.03577}

\bibitem[{{Cehula} {et~al.}(2024){Cehula}, {Thompson}, \& {Metzger}}]{2024MNRAS.528.5323C}
{Cehula}, J., {Thompson}, T.~A., \& {Metzger}, B.~D. 2024, \mnras, 528, 5323, \dodoi{10.1093/mnras/stae358}

\bibitem[{{Colaiuda} {et~al.}(2009){Colaiuda}, {Beyer}, \& {Kokkotas}}]{2009MNRAS.396.1441C}
{Colaiuda}, A., {Beyer}, H., \& {Kokkotas}, K.~D. 2009, MNRAS, 396, 1441, \dodoi{10.1111/j.1365-2966.2009.14878.x}

\bibitem[{{Cruise} {et~al.}(2025){Cruise}, {Guainazzi}, {Aird}, {Carrera}, {Costantini}, {Corrales}, {Dauser}, {Eckert}, {Gastaldello}, {Matsumoto}, {Osten}, {Petrucci}, {Porquet}, {Pratt}, {Rea}, {Reiprich}, {Simionescu}, {Spiga}, \& {Troja}}]{2025NatAs...9...36C}
{Cruise}, M., {Guainazzi}, M., {Aird}, J., {et~al.} 2025, Nature Astronomy, 9, 36, \dodoi{10.1038/s41550-024-02416-3}

\bibitem[{{Dillmann} {et~al.}(2025){Dillmann}, {Mart{\'\i}nez-Galarza}, {Soria}, {Stefano}, \& {Kashyap}}]{2025MNRAS.537..931D}
{Dillmann}, S., {Mart{\'\i}nez-Galarza}, J.~R., {Soria}, R., {Stefano}, R.~D., \& {Kashyap}, V.~L. 2025, \mnras, 537, 931, \dodoi{10.1093/mnras/stae2808}

\bibitem[{Feroci {et~al.}(2001)Feroci, Hurley, Duncan, \& Thompson}]{Feroci_2001}
Feroci, M., Hurley, K., Duncan, R.~C., \& Thompson, C. 2001, ApJ, 549, 1021, \dodoi{10.1086/319441}

\bibitem[{Frederiks {et~al.}(2007)Frederiks, Golenetskii, Palshin, Aptekar, Ilyinskii, Oleinik, Mazets, \& Cline}]{frederiks2007giant}
Frederiks, D., Golenetskii, S., Palshin, V., {et~al.} 2007, Astronomy Letters, 33, 1

\bibitem[{{Frederiks} {et~al.}(2007){Frederiks}, {Golenetskii}, {Palshin}, {Aptekar}, {Ilyinskii}, {Oleinik}, {Mazets}, \& {Cline}}]{2007AstL...33....1F}
{Frederiks}, D.~D., {Golenetskii}, S.~V., {Palshin}, V.~D., {et~al.} 2007, Astronomy Letters, 33, 1, \dodoi{10.1134/S106377370701001X}

\bibitem[{{Gehrels} {et~al.}(2004){Gehrels}, {Chincarini}, {Giommi}, {Mason}, {Nousek}, {Wells}, {White}, {Barthelmy}, {Burrows}, {Cominsky}, {Hurley}, {Marshall}, {M{\'e}sz{\'a}ros}, {Roming}, {Angelini}, {Barbier}, {Belloni}, {Campana}, {Caraveo}, {Chester}, {Citterio}, {Cline}, {Cropper}, {Cummings}, {Dean}, {Feigelson}, {Fenimore}, {Frail}, {Fruchter}, {Garmire}, {Gendreau}, {Ghisellini}, {Greiner}, {Hill}, {Hunsberger}, {Krimm}, {Kulkarni}, {Kumar}, {Lebrun}, {Lloyd-Ronning}, {Markwardt}, {Mattson}, {Mushotzky}, {Norris}, {Osborne}, {Paczynski}, {Palmer}, {Park}, {Parsons}, {Paul}, {Rees}, {Reynolds}, {Rhoads}, {Sasseen}, {Schaefer}, {Short}, {Smale}, {Smith}, {Stella}, {Tagliaferri}, {Takahashi}, {Tashiro}, {Townsley}, {Tueller}, {Turner}, {Vietri}, {Voges}, {Ward}, {Willingale}, {Zerbi}, \& {Zhang}}]{2004ApJ...611.1005G}
{Gehrels}, N., {Chincarini}, G., {Giommi}, P., {et~al.} 2004, \apj, 611, 1005, \dodoi{10.1086/422091}

\bibitem[{{Gelfand} {et~al.}(2005){Gelfand}, {Lyubarsky}, {Eichler}, {Gaensler}, {Taylor}, {Granot}, {Newton-McGee}, {Ramirez-Ruiz}, {Kouveliotou}, \& {Wijers}}]{Gelfand+05}
{Gelfand}, J.~D., {Lyubarsky}, Y.~E., {Eichler}, D., {et~al.} 2005, ApJ Letters, 634, L89, \dodoi{10.1086/498643}

\bibitem[{{Granot} {et~al.}(2006){Granot}, {Ramirez-Ruiz}, {Taylor}, {Eichler}, {Lyubarsky}, {Wijers}, {Gaensler}, {Gelfand}, \& {Kouveliotou}}]{Granot+06}
{Granot}, J., {Ramirez-Ruiz}, E., {Taylor}, G.~B., {et~al.} 2006, ApJ, 638, 391, \dodoi{10.1086/497680}

\bibitem[{{Hu} {et~al.}(2025){Hu}, {Wadiasingh}, {Ho}, {Baring}, {Younes}, {Enoto}, {Guillot}, {G{\"u}ver}, {Bause}, {Stewart}, {Van Kooten}, \& {Kouveliotou}}]{2025ApJ...989...63H}
{Hu}, C.-P., {Wadiasingh}, Z., {Ho}, W. C.~G., {et~al.} 2025, \apj, 989, 63, \dodoi{10.3847/1538-4357/adea4e}

\bibitem[{{Hurley} {et~al.}(2005){Hurley}, {Boggs}, {Smith}, {Duncan}, {Lin}, {Zoglauer}, {Krucker}, {Hurford}, {Hudson}, {Wigger}, {Hajdas}, {Thompson}, {Mitrofanov}, {Sanin}, {Boynton}, {Fellows}, {von Kienlin}, {Lichti}, {Rau}, \& {Cline}}]{Hurley+05}
{Hurley}, K., {Boggs}, S.~E., {Smith}, D.~M., {et~al.} 2005, Nature, 434, 1098, \dodoi{10.1038/nature03519}

\bibitem[{{Inan} {et~al.}(1999){Inan}, {Lehtinen}, {Lev-Tov}, {Johnson}, {Bell}, \& {Hurley}}]{1999GeoRL..26.3357I}
{Inan}, U.~S., {Lehtinen}, N.~G., {Lev-Tov}, S.~J., {et~al.} 1999, \grl, 26, 3357, \dodoi{10.1029/1999GL010690}

\bibitem[{{Kouveliotou} {et~al.}(1998){Kouveliotou}, {Dieters}, {Strohmayer}, {van Paradijs}, {Fishman}, {Meegan}, {Hurley}, {Kommers}, {Smith}, {Frail}, \& {Murakami}}]{kouveliotou1998x}
{Kouveliotou}, C., {Dieters}, S., {Strohmayer}, T., {et~al.} 1998, Nature, 393, 235, \dodoi{10.1038/30410}

\bibitem[{{Kouveliotou} {et~al.}(1999){Kouveliotou}, {Strohmayer}, {Hurley}, {van Paradijs}, {Finger}, {Dieters}, {Woods}, {Thompson}, \& {Duncan}}]{kouveliotou1998discovery}
{Kouveliotou}, C., {Strohmayer}, T., {Hurley}, K., {et~al.} 1999, ApJ Letters, 510, L115, \dodoi{10.1086/311813}

\bibitem[{{Lander}(2016)}]{2016ApJ...824L..21L}
{Lander}, S.~K. 2016, \apjl, 824, L21, \dodoi{10.3847/2041-8205/824/2/L21}

\bibitem[{{Lander}(2023)}]{2023ApJ...947L..16L}
---. 2023, ApJ Letters, 947, L16, \dodoi{10.3847/2041-8213/acca1f}

\bibitem[{{Lander} {et~al.}(2015){Lander}, {Andersson}, {Antonopoulou}, \& {Watts}}]{2015MNRAS.449.2047L}
{Lander}, S.~K., {Andersson}, N., {Antonopoulou}, D., \& {Watts}, A.~L. 2015, MNRAS, 449, 2047, \dodoi{10.1093/mnras/stv432}

\bibitem[{{Levin}(2006)}]{2006MNRAS.368L..35L}
{Levin}, Y. 2006, MNRAS, 368, L35, \dodoi{10.1111/j.1745-3933.2006.00155.x}

\bibitem[{{Liang} \& {Antiochos}(1984)}]{1984Natur.310..121L}
{Liang}, E.~P., \& {Antiochos}, S.~K. 1984, \nat, 310, 121, \dodoi{10.1038/310121a0}

\bibitem[{{Mereghetti}(2008)}]{2008AandARv..15..225M}
{Mereghetti}, S. 2008, A\&A Review, 15, 225, \dodoi{10.1007/s00159-008-0011-z}

\bibitem[{{Mereghetti} {et~al.}(2005){Mereghetti}, {G{\"o}tz}, {von Kienlin}, {Rau}, {Lichti}, {Weidenspointner}, \& {Jean}}]{2005ApJ...624L.105M}
{Mereghetti}, S., {G{\"o}tz}, D., {von Kienlin}, A., {et~al.} 2005, \apjl, 624, L105, \dodoi{10.1086/430669}

\bibitem[{{Mereghetti} {et~al.}(2015){Mereghetti}, {Pons}, \& {Melatos}}]{2015SSRv..191..315M}
{Mereghetti}, S., {Pons}, J.~A., \& {Melatos}, A. 2015, Space Science Reviews, 191, 315, \dodoi{10.1007/s11214-015-0146-y}

\bibitem[{{Mereghetti} {et~al.}(2024){Mereghetti}, {Rigoselli}, {Salvaterra}, {Pacholski}, {Rodi}, {Gotz}, {Arrigoni}, {D'Avanzo}, {Adami}, {Bazzano}, {Bozzo}, {Brivio}, {Campana}, {Cappellaro}, {Chenevez}, {De Luise}, {Ducci}, {Esposito}, {Ferrigno}, {Ferro}, {Israel}, {Le Floc'h}, {Martin-Carrillo}, {Onori}, {Rea}, {Reguitti}, {Savchenko}, {Souami}, {Tartaglia}, {Thuillot}, {Tiengo}, {Tomasella}, {Topinka}, {Turpin}, \& {Ubertini}}]{2024Natur.629...58M}
{Mereghetti}, S., {Rigoselli}, M., {Salvaterra}, R., {et~al.} 2024, \nat, 629, 58, \dodoi{10.1038/s41586-024-07285-4}

\bibitem[{{Negro} {et~al.}(2024){Negro}, {Younes}, {Wadiasingh}, {Burns}, {Trigg}, \& {Baring}}]{2024FrASS..1188953N}
{Negro}, M., {Younes}, G., {Wadiasingh}, Z., {et~al.} 2024, Frontiers in Astronomy and Space Sciences, 11, 1388953, \dodoi{10.3389/fspas.2024.1388953}

\bibitem[{{Patel} {et~al.}(2025{\natexlab{a}}){Patel}, {Metzger}, {Cehula}, {Burns}, {Goldberg}, \& {Thompson}}]{2025ApJ...984L..29P}
{Patel}, A., {Metzger}, B.~D., {Cehula}, J., {et~al.} 2025{\natexlab{a}}, \apjl, 984, L29, \dodoi{10.3847/2041-8213/adc9b0}

\bibitem[{{Patel} {et~al.}(2025{\natexlab{b}}){Patel}, {Metzger}, {Goldberg}, {Cehula}, {Thompson}, \& {Renzo}}]{Patel+2025b}
{Patel}, A., {Metzger}, B.~D., {Goldberg}, J.~A., {et~al.} 2025{\natexlab{b}}, \apj, 985, 234, \dodoi{10.3847/1538-4357/adceb7}

\bibitem[{{Pearlman} {et~al.}(2025){Pearlman}, {Scholz}, {Bethapudi}, {Hessels}, {Kaspi}, {Kirsten}, {Nimmo}, {Spitler}, {Fonseca}, {Meyers}, {Stairs}, {Tan}, {Bhardwaj}, {Chatterjee}, {Cook}, {Curtin}, {Dong}, {Eftekhari}, {Gaensler}, {G{\"u}ver}, {Kaczmarek}, {Leung}, {Masui}, {Michilli}, {Prince}, {Sand}, {Shin}, {Smith}, \& {Tendulkar}}]{2025NatAs...9..111P}
{Pearlman}, A.~B., {Scholz}, P., {Bethapudi}, S., {et~al.} 2025, Nature Astronomy, 9, 111, \dodoi{10.1038/s41550-024-02386-6}

\bibitem[{{Quirola-V{\'a}squez} {et~al.}(2022){Quirola-V{\'a}squez}, {Bauer}, {Jonker}, {Brandt}, {Yang}, {Levan}, {Xue}, {Eappachen}, {Zheng}, \& {Luo}}]{2022A&A...663A.168Q}
{Quirola-V{\'a}squez}, J., {Bauer}, F.~E., {Jonker}, P.~G., {et~al.} 2022, \aap, 663, A168, \dodoi{10.1051/0004-6361/202243047}

\bibitem[{{Quirola-V{\'a}squez} {et~al.}(2023){Quirola-V{\'a}squez}, {Bauer}, {Jonker}, {Brandt}, {Yang}, {Levan}, {Xue}, {Eappachen}, {Camacho}, {Ravasio}, {Zheng}, \& {Luo}}]{2023A&A...675A..44Q}
---. 2023, \aap, 675, A44, \dodoi{10.1051/0004-6361/202345912}

\bibitem[{{Rea} {et~al.}(2013){Rea}, {Esposito}, {Pons}, {Turolla}, {Torres}, {Israel}, {Possenti}, {Burgay}, {Vigan{\`o}}, {Papitto}, {Perna}, {Stella}, {Ponti}, {Baganoff}, {Haggard}, {Camero-Arranz}, {Zane}, {Minter}, {Mereghetti}, {Tiengo}, {Sch{\"o}del}, {Feroci}, {Mignani}, \& {G{\"o}tz}}]{2013ApJ...775L..34R}
{Rea}, N., {Esposito}, P., {Pons}, J.~A., {et~al.} 2013, ApJ Letters, 775, L34, \dodoi{10.1088/2041-8205/775/2/L34}

\bibitem[{{Reynolds} {et~al.}(2023){Reynolds}, {Kara}, {Mushotzky}, {Ptak}, {Koss}, {Williams}, {Allen}, {Bauer}, {Bautz}, {Bogadhee}, {Burdge}, {Cappelluti}, {Cenko}, {Chartas}, {Chan}, {Corrales}, {Daylan}, {Falcone}, {Foord}, {Grant}, {Habouzit}, {Haggard}, {Herrmann}, {Hodges-Kluck}, {Kargaltsev}, {King}, {Kounkel}, {Lopez}, {Marchesi}, {McDonald}, {Meyer}, {Miller}, {Nynka}, {Okajima}, {Pacucci}, {Russell}, {Safi-Harb}, {Strassun}, {Trindade Falc{\~a}o}, {Walker}, {Wilms}, {Yukita}, \& {Zhang}}]{2023SPIE12678E..1ER}
{Reynolds}, C.~S., {Kara}, E.~A., {Mushotzky}, R.~F., {et~al.} 2023, in Society of Photo-Optical Instrumentation Engineers (SPIE) Conference Series, Vol. 12678, UV, X-Ray, and Gamma-Ray Space Instrumentation for Astronomy XXIII, ed. O.~H. {Siegmund} \& K.~{Hoadley}, 126781E, \dodoi{10.1117/12.2677468}

\bibitem[{{Ripley} {et~al.}(2014){Ripley}, {Metzger}, {Arcones}, \& {Mart{\'\i}nez-Pinedo}}]{2014MNRAS.438.3243R}
{Ripley}, J.~L., {Metzger}, B.~D., {Arcones}, A., \& {Mart{\'\i}nez-Pinedo}, G. 2014, \mnras, 438, 3243, \dodoi{10.1093/mnras/stt2434}

\bibitem[{Roberts {et~al.}(2021)Roberts, Veres, Baring, Briggs, Kouveliotou, Bissaldi, Younes, Chastain, DeLaunay, Huppenkothen, {et~al.}}]{roberts2021rapid}
Roberts, O., Veres, P., Baring, M., {et~al.} 2021, Nature, 589, 207

\bibitem[{{Rodi} {et~al.}(2025){Rodi}, {Pacholski}, {Mereghetti}, {Arrigoni}, {Bazzano}, {Natalucci}, {Salvaterra}, \& {Ubertini}}]{2025ApJ...979L..25R}
{Rodi}, J.~C., {Pacholski}, D.~P., {Mereghetti}, S., {et~al.} 2025, \apjl, 979, L25, \dodoi{10.3847/2041-8213/ada6b7}

\bibitem[{{Safi-Harb} {et~al.}(2023){Safi-Harb}, {Burdge}, {Bodaghee}, {An}, {Guest}, {Hare}, {Hebbar}, {Ho}, {Kargaltsev}, {Kirmizibayrak}, {Klingler}, {Nynka}, {Reynolds}, {Sasaki}, {Sridhar}, {Vasilopoulos}, {Woods}, {Yang}, {Heinke}, {Kong}, {Li}, {MacMaster}, {Mallick}, {Treyturik}, {Tsuji}, {Binder}, {Braun}, {Chang}, {Chatterjee}, {Ferrand}, {Holland-Ashford}, {Ng}, {Plotkin}, {Romani}, \& {Zhang}}]{2023arXiv231107673S}
{Safi-Harb}, S., {Burdge}, K.~B., {Bodaghee}, A., {et~al.} 2023, arXiv e-prints, arXiv:2311.07673, \dodoi{10.48550/arXiv.2311.07673}

\bibitem[{{Thompson} \& {Duncan}(1995)}]{1995MNRAS.275..255T}
{Thompson}, C., \& {Duncan}, R.~C. 1995, \mnras, 275, 255, \dodoi{10.1093/mnras/275.2.255}

\bibitem[{Trigg {et~al.}(2023)Trigg, Burns, Roberts, Negro, Svinkin, Baring, Wadiasingh, Christensen, Andreoni, Briggs, {et~al.}}]{trigg2023grb}
Trigg, A.~C., Burns, E., Roberts, O.~J., {et~al.} 2023, arXiv preprint arXiv:2311.09362

\bibitem[{{Trigg} {et~al.}(2025){Trigg}, {Stewart}, {Van Kooten}, {Burns}, {Baring}, {Frederiks}, {Huppenkothen}, {O'Connor}, {Roberts}, {Wadiasingh}, {Younes}, {Bhat}, {Briggs}, {Busmann}, {Goldstein}, {Gruen}, {Hu}, {Kouveliotou}, {Negro}, {Palmese}, {Riffeser}, {Scotton}, {Svinkin}, {Veres}, \& {Z{\"o}ller}}]{2025A&A...694A.323T}
{Trigg}, A.~C., {Stewart}, R., {Van Kooten}, A., {et~al.} 2025, \aap, 694, A323, \dodoi{10.1051/0004-6361/202452268}

\bibitem[{{Watts} \& {Strohmayer}(2007)}]{watts2007neutron}
{Watts}, A.~L., \& {Strohmayer}, T.~E. 2007, Advances in Space Research, 40, 1446, \dodoi{10.1016/j.asr.2006.12.021}

\end{thebibliography}
\bibliographystyle{aasjournal}

\end{document}